\documentclass{aa}
\usepackage[utf8]{inputenc}
\usepackage{ngerman}
\usepackage{verbatim}
\usepackage{amsmath}
\usepackage{xcolor}
\usepackage{mathcomp}
\usepackage{float}
\usepackage{array}
\usepackage{graphicx}
\usepackage{geometry}
\usepackage{url}
\usepackage[hidelinks]{hyperref} 
\usepackage[separate-uncertainty]{siunitx}
\bibpunct{(}{)}{;}{a}{}{,}
\date{Received date /
Accepted date }
\begin{document}
\title{Thermal evolution of Uranus and Neptune I: adiabatic models}
\author{Ludwig Scheibe\inst{1} \and Nadine Nettelmann\inst{1,2} \and Ronald Redmer\inst{1}}
\institute{Institut für Physik, Universität Rostock, A.-Einstein-Str. 23, D-18059 Rostock, Germany \and Institut für Planetenforschung, Deutsches Zentrum für Luft- und Raumfahrt,  Rutherfordstraße 2, D-12489 Berlin, Germany}
\abstract{
The brightness of Neptune is often found to be in accordance with an adiabatic interior, while the low luminosity of Uranus challenges this assumption. Here we apply revised equation of state data of hydrogen, helium, and water and compute the thermal evolution of Uranus and Neptune assuming an adiabatic interior. For this purpose, we have developed a new planetary model and evolution code. We investigate the influence of albedo, solar energy influx, and equations of state of H and He, and water on the cooling time. Our cooling times of about $\tau_\text{U}=\SI{5.1e9}{years}$ for Uranus and $\tau_\text{N}=\SI{3.7e9}{years}$ for Neptune bracket the known age of the planets of $\SI{4.56e9}{years}$ implying that neither planet's present-day luminosity can be explained by adiabatic cooling. We also find that uncertainties on input parameters such as the level of irradiation matter generally more for Uranus than for Neptune. Our results suggest that  in contrast to common assumptions, neither planet is fully adiabatic in the deeper interior.
}
\titlerunning{Thermal evolution of Uranus and Neptune I}
\maketitle
\section{Introduction}
Uranus and Neptune, the two outermost planets of our solar system, share a large number of very similar observed values such as mean density, surface temperature, atmospheric composition, and  magnetic field morphology \citep{Guillot15}. These observations suggest similar  structural and evolutionary paths since their time of formation $\SI{4.56e9}{years}$ ago. Because of their similar characteristics, which are different from the larger primarily hydrogen- and helium-based gas giants Jupiter and Saturn and the smaller rock-based inner planets, they are usually classified in their own category as ice giants. \\ 
Neptune's intrinsic heat flux of $F_\text{int, N} = \SI{0.433 \pm 0.046}{\watt \metre ^{-2}}$ is about an order of magnitude higher than that of Uranus ($F_\text{int,U}=0.042_{-0.042}^{+0.047}\: \si{\watt \metre ^{-2}}$) \citep{Guillot15}.  Uranus' intrinsic flux is consistent with being zero, meaning that it could be in thermal equilibrium with the Sun, while Neptune is clearly still cooling. \\
Earlier adiabatic cooling calculations using a zero-temperature equation of state (EOS) for the ice material have found that both planets  are cooling too slowly to explain their present-day luminosity \citep{Hubbard78, Hubbard80, Hubbard95}. More recent models, using a finite-temperature EOS for water but currently outdated ones for hydrogen and helium, could reproduce Neptune's brightness but still obtained too high luminosity for Uranus \citep{Fortney11, Nettelmann13, Linder19}.\\
It is important to gain better insight into the thermal structure of Uranus and Neptune and their heat fluxes because the thermal structure largely influences the inferred composition \citep{Podolak19}. As of June 2019, out of the 3972 confirmed and categorised exoplanets\footnote{\url{https://exoplanets.nasa.gov/}}, more than 30\% are in a size range similar to Neptune and about 30\% are intermediate in size between Earth and Neptune. Due to selection effects in exoplanet surveys, however, these numbers refer to the detected planets, while the absolute occurrence rates at large orbital distances (>100 days) are not known yet. Nevertheless, understanding Uranus and Neptune is an important step towards understanding a frequently occurring class of exoplanets.\\ 
In this work, we apply the Rostock Equation of State version 3 \citep[REOS.3]{Becker14} for hydrogen and helium and the recently published water EOS by \citet{Mazevet19} to calculate new adiabatic cooling times for Uranus and Neptune.  We investigate the influence of solar energy influx and EOS data on the cooling behaviour. To this end, we present the theoretical foundations of a newly developed computer code in sect.~\ref{sec:Methods} and calculate a variety of planetary cooling tracks with different assumptions. The results are presented and discussed in sect.~\ref{sec:results}. Section~\ref{sec:conclusion} summarises the results and  gives a brief outlook. 
\section{Methodology}
\label{sec:Methods}
\subsection{Theoretical foundations}
To compute the thermal evolution of fluid planets we  developed a new tool, which we  named \emph{OTher Thermal Evolution Realisation} (OTTER). It is based on the following basic equations, representing conservation of mass, hydrostatic equilibrium, energy transport, and conservation of energy for the main variables of radius $r(m)$, pressure $P(m)$, temperature $T(m)$, and luminosity $l(m)$ \mbox{\citep{KippWei}}:
\begin{align}
	f_0 &= \dfrac{\partial  r}{\partial m} = \frac{1}{4 \pi\, r^2\, \rho}, \label{eq:rm}\\
        f_1 &= \dfrac{\partial  P}{\partial m} = -\frac{G\,m}{4\pi\, r^4} + \frac{\omega^2}{6\pi\, r\,}, \label{eq:Pm}\\
        f_2 &= \dfrac{\partial  T}{\partial m} = \left(  -\frac{G\,m}{4\pi\, r^4} + \frac{\omega^2}{6\pi\, r } \right) \frac{T}{P}\nabla_{T}, \label{eq:Tm}\\
        f_3 &= \dfrac{\partial  l}{\partial m} = - T \dfrac{\partial  s}{\partial t} = - \dfrac{\partial  u}{\partial t} + \frac{P}{\rho^2} \dfrac{\partial  \rho}{\partial t} = - c_p \dfrac{\partial  T}{\partial t} + \frac{\delta}{\rho} \dfrac{\partial  P}{\partial t}.\label{eq:lm}
\end{align}
Equations \eqref{eq:Pm} and \eqref{eq:Tm} contain terms arising from rigid body rotation in zeroth order. In equation \eqref{eq:Tm}, $\nabla_T=\dfrac{\partial  \ln T}{\partial \ln P}$ denotes the temperature gradient that is evaluated at each point separately. In this work we follow the commonly used assumption of a convective adiabatic interior for gas and ice giants \citep{Guillot99, Helled11}, which means   for $\nabla_T$ we use the adiabatic gradient \citep{KippWei}
\begin{equation}
  \nabla_{T,\text{ad}}=\left( \dfrac{\partial  \ln T}{\partial \ln P} \right)_S=\frac{P\: \delta}{T\:\rho\: c_p},
\end{equation}
where $\delta=-\left( \dfrac{\partial  \ln\rho}{\partial \ln T} \right)_P$ is the thermal expansion coefficient and $c_p=\left( \dfrac{\partial  u}{\partial T} \right)_P - \frac{P}{\rho^2} \left( \dfrac{\partial  \rho}{\partial T} \right)_P$ the isobaric heat capacity. \\
The two outer boundary conditions at $m=M_\text{P}$ are 
\begin{align}
  P (M_\text{P})&=\SI{1}{\bar}, \label{eq:B0}\\
  l (M_\text{P})&=4 \pi R^2 \sigma_\text{B} T_\text{eff}^4 - L_\text{sol}, \label{eq:B1}
\end{align}
where $\sigma_\text{B}=\SI{5.6704e-8}{W m^{-2} K^{-4}}$ is the Stefan-Boltzmann constant, $L_\text{sol}$ the solar radiation absorbed and re-emitted by the planet, and $T_\text{eff}$ denotes the planet's effective temperature. We link it to the one-bar temperature via 
\begin{equation}
  T_\text{1bar}=K g^{-1/6} T_\text{eff}^{1.244}. \label{eq:1bar_eff}
\end{equation}
Equation \eqref{eq:1bar_eff} is based on model atmospheres for Jupiter developed by \citet{Grab75} and interpolated by \citet{Hubb77}. The parameter $K$ is chosen to yield the present-day one-bar
temperatures for Uranus and Neptune for their present-day radius and effective temperature. We find $K_\text{U}=1.481$ for $T_\text{1bar}=\SI{76}{\kelvin}$ for Uranus and $K_\text{N}=1.451$ for $T_\text{1bar}=\SI{72}{\kelvin}$ for Neptune. Compared to the model atmosphere by \citet{Fortney11} developed for Uranus and Neptune, we find that $K$ changes only by a small percent with $T_\text{eff}$ and $g$ except for high values of $T_\text{eff}$, where the cooling time is short. We therefore assume it to be constant over the planet's lifetime.\\
Following \citet{KippWei}, we define four functions $G_i^j$ at the $j$th point of a mass grid for a given time
\begin{align}
  &G_i^j := \frac{y_i^j - y_i^{j+1}}{\xi^j - \xi^{j+1}} - f_i\left(y_0^{j+\frac{1}{2}}, y_1^{j+\frac{1}{2}}, y_2^{j+\frac{1}{2}}, y_3^{j+\frac{1}{2}} \right),\\
  &i=0, \ldots,3. \nonumber
\end{align}
Here $y_i$ are the logarithmic main variables $y_0=\ln (r), y_1=\ln (P), y_2=\ln (T), y_3=\ln (l)$; the values $y_i^{j+1/2}$ are obtained as the arithmetic mean of the values at the $j$th and $(j+1)$st point. The functions $f_i$ are the derivatives of the main variables as defined in equations \eqref{eq:rm}-\eqref{eq:lm}. We use $\xi=\ln \left( 1 - \frac{m}{1.05\: M_\text{p}} \right)$ as our mass coordinate, following \citet{Kippenhahn67}. Approximating the derivatives as the corresponding difference coefficient, a profile that fulfils equation \eqref{eq:rm} - \eqref{eq:lm} must satisfy $G_i^j=0$. Similarly, the two outer boundary conditions \eqref{eq:B0} and \eqref{eq:B1} are formulated as $B_0$ and $B_1$. Therefore, for $N$ mass points, we have a system of $4N-2$ equations. Using Newton's method, it can be solved by recasting to the matrix equation
\begin{center}
\begin{equation}
  \underbrace{
  \begin{pmatrix}
    \dfrac{\partial  B_0}{\partial y_0^0} & & \cdots  \\
     & \ddots & \\
    \cdots & & \dfrac{\partial  G_{N-1}^3}{\partial y_2^{N-1}}
  \end{pmatrix}
  }_{H}
  \cdot
  \underbrace{
  \begin{pmatrix}
    \delta y_0^0 \vphantom{\dfrac{\partial  B_0}{\partial y_0^0}} \\
    \vdots \\
    \delta y_2^{N-1} \vphantom{\dfrac{\partial  G_{N-1}^3}{\partial y_2^{N-1}}}
  \end{pmatrix}
  }_{\vec{\delta y}}
  = - 
  \underbrace{
  \begin{pmatrix}
    B_0 \vphantom{\dfrac{\partial  B_0}{\partial y_0^0}}\\
    \vdots \\
    G_{N-1}^3 \vphantom{\dfrac{\partial  G_{N-1}^3}{\partial y_2^{N-1}}}
  \end{pmatrix}
  }_{\vec{G}} .
\end{equation}
\end{center}
For a given time, the matrix $H$ and the vector $\vec{G}$ are constructed using an estimated  solution. Solving the equation above then yields  corrections $\delta y_i^j$ for every main variable at every mass point, which are added to the estimate to generate a new input. This is repeated until all the corrections fulfil the truncation condition $\delta y_i^j < \num{5e-8}$, which means the profile is converged. The difference between using this condition in comparison to $\delta y_i^j < \num{1e-10}$ lies in the tenth digit of the effective cooling time.
Then, a new estimated solution for the next time step is estimated based on the current and the previous time step, and a new profile is calculated for the new time. \\
The planet's rotation period is fixed over the whole evolution time at $t_{\omega,\text{U}}=\SI{6.206e4}{s}$ for Uranus and $t_{\omega,\text{N}}=\SI{5.8e4}{s}$ for Neptune \citep{Guillot15}. Accounting for angular momentum conservation leads to a change in rotational period of about ten hours over the course of the planet's evolution; however, we found only a small effect on the resulting cooling times and therefore neglect it for numerical simplicity. 
The program is coded in C++ and uses sparse matrix functions from the Eigen-library \citep{Eigen} to solve the relevant matrix equations. 
\subsection{Equations of state}
\label{sec:eos}
Uranus and Neptune are commonly assumed to be composed primarily of conducting fluid ices as well as hydrogen and helium \citep{Podolak95}. \\
In this paper we apply four different EOSs for hydrogen and helium. For benchmark purposes,  in sect.~\ref{sec:benchmark} we use the SCvH H/He EOS \citep{SCvH95}, while in sect.~\ref{sec:results}, for our adiabatic Uranus and Neptune models, we apply  the two versions of the Rostock EOS, REOS.1 \citep{Nettelmann08} and REOS.3 \citep{Becker14}, and  the recent EOS by \citet{Chabrier19}. \\
In H-REOS.3, the DFT-MD data are based on simulations using 256 particles (H-REOS.1: 64 particles). In H-REOS.3, the internal energy is corrected for the quantum mechanical energy of harmonic oscillations  of the H$_2$ molecule, while H-REOS.1 only accounts for the classical treatment of the vibrations. These two changes were already applied to H-REOS.2 \citep{Nettelmann12}; however, that EOS was built on a coarser temperature grid at low temperatures subject to larger uncertainties upon interpolation. \\
He-REOS.3 differs from He-REOS.1 in that it is based on DFT-MD data using 108 particles (He-REOS: 32-64) on a significantly denser temperature grid than He-REOS.1. Moreover, the DFT-MD data in He-REOS.3 are connected to virial EOS data at $T-\rho$ conditions relevant to planets,  namely at $\rho < ~\SI{0.1}{\gram\per\cubic\centi\metre}$ and $T < \SI{1e4}{\kelvin}$, while at high temperatures more relevant to hot Jupiters and brown dwarfs it connects to the He-SCvH EOS. Instead,  He-REOS.1 exclusively connects to Sesame EOS 5761 at low densities and/or high temperatures.\\
The EOS by \citet{Chabrier19} (hereafter  Chabrier-EOS) uses the SCvH data in the low-temperature and low-pressure regime.  At high temperatures and high pressures, where the material is fully ionised, it utilises data from \citet{Chabrier98}. In the intermediate regime of partial ionisation and dissociation, the Chabrier-EOS employs results of quantum molecular dynamics calculations. The different data sets are combined via a bicubic spline interpolation. Like the SCvH data, this EOS provides the entropy and adiabatic temperature gradient explicitly. Since it does not extend to $T<\SI{100}{K}$, models featuring this EOS use REOS.3-data for H and He for low temperatures.\\
In the representation of ices, here we use  EOSs of pure water as in \citet{Helled11} and \citet{Nettelmann13}. This is certainly a simplification since the further ice-forming elements carbon, nitrogen, and sulphur are likely to be present in the interior as well. Former work has shown that the presence of ices lighter than water in solar element ratios can lead to the extreme case of an icy inner envelope \citep{Podolak87} and hence a cooler planet \citep{Bethkenhagen17}. On the other hand, rocks may be soluble in hydrogen and thus be present to some degree in the envelope at the expense of ices, while methane may be insoluble in water and may lead to carbon sedimentation. Given these uncertainties on the deep interior composition, we believe that the uncertainty added by our simplification does not change the conclusions of this work.\\
We consider two water equations of state. The Sesame 7150 EOS \citep{Ree76} is a combination of experimental work and chemical model data connected via interpolation, which makes it quite accurate for lower temperatures but less so for the warm dense interior of the ice giants where there are few experimental results (see \citet{Knudson12} for a comparison of water EOS data along the principal Hugoniot).  Recently, a new water EOS was released by \citet{Mazevet19}. It is based on a thermodynamically consistent Helmholtz free energy parameterisation over a wide range of temperature-pressure conditions and incorporates EOS data from various sources, including ab initio simulation data. In order to handle the low-temperature phase transitions between solid, liquid, and gaseous water below the critical point, we  applied a Maxwell construction on the water EOS of \citet{Mazevet19}, in the following labelled ``Mazevet EOS''.   However, due to numerical difficulties in handling first-order phase transition, we use an ideal gas description for water for $T < \SI{800}{K}$, independently of whether the Sesame-EOS or Mazevet-EOS is used for the majority of the planet. This ideal gas description contains quantum mechanical corrections of the vibrational and rotational energy of the H$_2$O-molecule following \citet{French09}. \\
The different  EOS data of different materials are combined via the linear mixing rule, $\rho^{-1}=\sum_i X_i/\rho_i$ and $u=\sum_iX_i u_i$, where $X_i$ are the different species' mass fractions, which is a reasonable assumption, as shown by \citet{Bethkenhagen17}. The rock core is represented by the Sesame 7530 EOS for basalt \citep{Lyon92}. 
\subsection{Benchmarking with respect to MOGROP}
\label{sec:benchmark}
\begin{figure}
  \resizebox{\hsize}{!}{\includegraphics{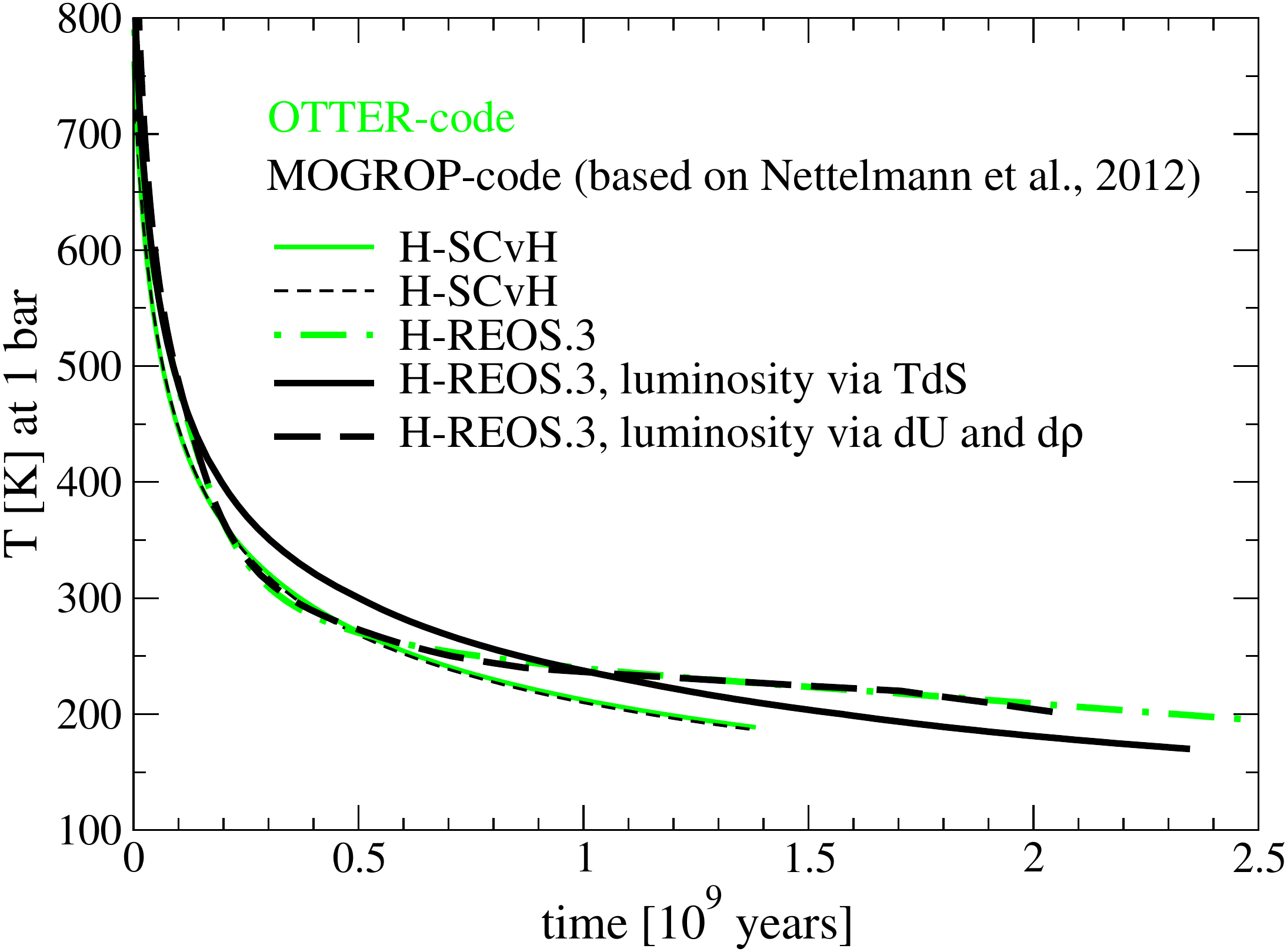}}
  \caption{Evolution of a $300 M_\text{E}$ planet with a $2 M_\text{E}$ rock core and an adiabatic hydrogen envelope. Shown are comparisons between the OTTER code (red, this work) and the MOGROP code (black, \citet{Nettelmann12}) for two different hydrogen equations of state--thin: SCvH-EOS \citep{SCvH95}; thick: REOS.3 \citep{Becker14}--and, for MOGROP, different ways of calculating the luminosity.} \label{fig:300Me}
\end{figure}

To verify the results provided by our new code, we compared our evolution curves obtained with OTTER to evolution curves using the MOGROP code \citep{Nettelmann12}. For this purpose, a test planet of $M_\text{P}=300 M_\text{E}$ with a small rock core and an adiabatic hydrogen envelope was considered.\\
Figure~\ref{fig:300Me} shows that both codes produce similar thermal evolution tracks if the hydrogen EOS by \citet{SCvH95} is used for the envelope, which provides explicitly the adiabatic temperature gradient and the entropy. However, when using the H-REOS.3 hydrogen EOS table \citep{Becker14}, which does not include these quantities, the level of agreement depends on the method of calculating the luminosity. Using MOGROP's standard way of calculating the heat, which draws on the planet's entropy profile calculated via thermodynamic integration from internal energy and density, we find a large difference from OTTER's evolution curve (compare the black solid and the green dot-dashed curves in Fig.~\ref{fig:300Me}). However, using interior energy and density directly instead of entropy in MOGROP (cf.\ equation \eqref{eq:lm}), we find good agreement with the new code (compare the black dashed and the red dot-dashed curves in Fig.~\ref{fig:300Me}). As the ultimate reason for this behaviour, we identify imperfect thermodynamic consistency in the H-REOS-table which leads to amplified uncertainties on the entropy,  which in the MOGROP code is derived by thermodynamic integration on the EOS data $u(T, \rho), P(T, \rho)$ \citep[see][]{Nettelmann12}. The uncertainties on  the heat $\delta Q$ are apparently much smaller when simply interpolating in these data directly. \\
These findings suggest that there may be larger uncertainties on previous evolution calculations than assumed. It also emphasises the need for wide-range thermodynamically consistent EOSs in planetary modelling for the most abundant materials H, He, and H$_2$O. 
\section{Results}
\label{sec:results}
\subsection{Interior models}
Below we present our results for interior models and thermal evolution curves for Uranus and Neptune. The models employ an adiabatic temperature profile and are separated into a hydrogen- and helium-rich outer envelope with a certain amount of water as a representation of heavy elements, a water-rich inner envelope with only a small fraction of hydrogen and helium, and a small isothermal rock core. It has been shown by \citet{Linder19} that there is little difference in luminosity between an isothermal and an adiabatic rock core. The structure models have a discontinuity in composition at about $\SI{30}{GPa}$ between the outer and inner envelope.  Table~\ref{tab:parameters} shows the parameters of the structure models, which were chosen to reproduce the planet's present day mean radius. For each planet there are two structure models, one using the Sesame 7150 EOS for water \citep{Ree76}, the other using the EOS by \citet{Mazevet19} (see sect.~\ref{sec:eos} for details).\\
The total mass of hydrogen and helium also given in the table is consistent with values of previously published models that satisfy the gravity data \citep{HelNetGui19}. However, hydrogen and helium in our models are more concentrated towards the surface. Consequently, the gravitational moments $J_2, J_4$ are lower than the measured values \citep[cf.][]{Nettelmann13} by about $\sim \SI{20}{\%}$ for Uranus and $\sim \SI{5}{\%}$ for Neptune. Due to the sensitivity of Uranus' cooling times to small changes in the assumptions (see sect.~\ref{sec:sol} for details), this could lead to a shift in Uranus' effective cooling time of up to $\SI{1}{Gyr}$. However, the systematic behaviour with the parameters studied here remains unaffected by this.\\
In the following segments, we investigate the effect of  different factors on the cooling behaviour of adiabatic Uranus and Neptune. In all figures corresponding to the various evolution calculations, the curves are shifted to run through the present-day $T_\text{eff}$ for $t=4.56\cdot 10^9$ years, the age of the solar system, to allow for an easy comparison of the differences in cooling times at first glance.\\
\begin{table*}
  \caption{Model parameters for the Uranus and Neptune structure models, as well as EOS data used for water (cf.\ sect.~\ref{sec:eos} for details). The quantities are outer envelope metallicity $Z_1$, inner envelope metallicity $Z_2$, transition mass between outer and inner fluid envelope $M_{12}$, core mass $M_\text{core}$, and total mass of combined hydrogen and helium $M_\text{H,He}$.} 
  \centering
  \begin{tabular}{l l l c c c c c }
	\hline \hline
    	 & H,He-EOS &H$_2$O-EOS & $Z_1$ & $Z_2$& $M_{12}$& $M_\text{core}$ & $M_\text{H,He}$\\
	 & & & & &  [$M_\text{E}$] &  [$M_\text{E}$] & [$M_\text{E}$] \\
    	\hline 
	Uranus 1 & REOS.3 & Sesame & 0.261 & 0.941& 12.5 & 0.25 & 2.2 \\
	Uranus 2 & REOS.3 & Mazevet & 0.273 & 0.96 & 12.44 & 0.79 & 2.0\\
	Neptune 1 &REOS.3 & Sesame & 0.42 & 0.91&	15.0 & 1.06 & 2.4\\
	Neptune 2 & REOS.3 & Mazevet & 0.42 & 0.929 & 15.17 & 1.04 & 2.2\\
	Neptune 3 & Chabrier & Mazevet & 0.424 & 0.92 & 15.17 & 1.05 & 2.3\\
	\hline
  \end{tabular}\label{tab:parameters}
\end{table*}
\subsection{Influence of hydrogen and helium equation of state}
\begin{figure}
  \resizebox{\hsize}{!}{\includegraphics{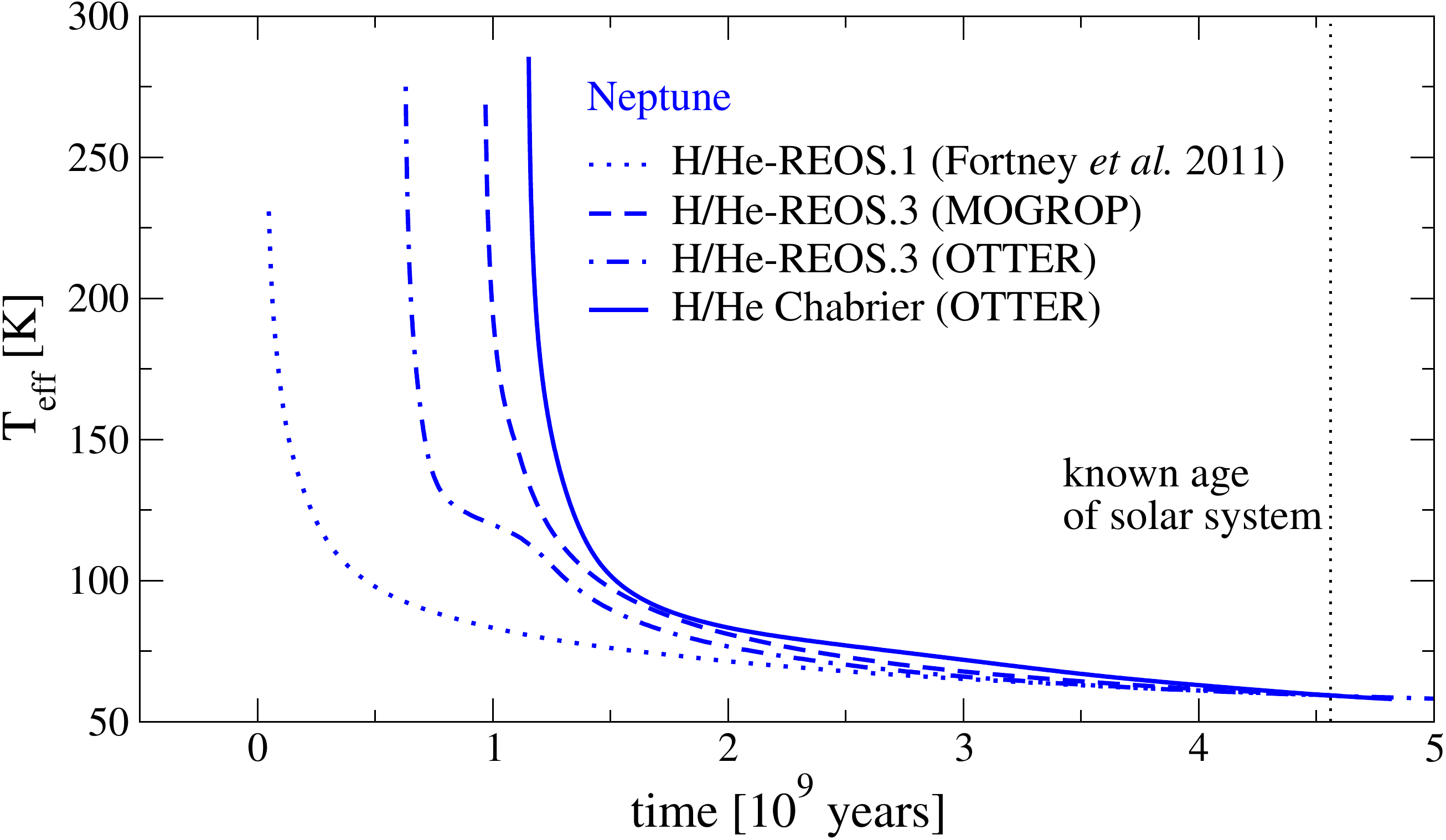}}
  \caption{Evolution calculations for Neptune using different H/He EOS data and codes: dotted: H/He-REOS.1, MOGROP \citep[adapted from][]{Fortney11}; dashed: H/He-REOS.3, MOGROP; dash-dotted: H/He-REOS.3, OTTER; solid: H/He-Chabrier, OTTER. All curves use $T_\text{eq}=\text{const.}$ (see sect.~\ref{sec:sol}). }  \label{fig:HHe}
\end{figure}

While Uranus and Neptune are thought to be composed primarily of higher density ices like water, ammonia, and methane, their gravitational moments cannot be reproduced without the presence of the lighter elements hydrogen and helium \citep{Hubbard80, Podolak87, Helled11, Podolak19, Nettelmann13}. 
In fig.~\ref{fig:HHe} we investigate the influence of the  H/He EOS (for details, see sect.~\ref{sec:eos}). We find that using the REOS.3 table for H and He yields a shortening of cooling times for Neptune by at least \SI{0.6}{Gyr} compared to previous work. The exact amount depends somewhat on the evolution code used and on the water EOS (see sect.~\ref{sec:H2OEOS}). Nevertheless, the shortening is visible with both MOGROP and OTTER and seems to be caused by the different H/He EOS. Using the Chabrier-EOS for H/He leads to a further shortening of Neptune's cooling time by about \SI{0.5}{Gyr}, underscoring the importance of H and He for the cooling behaviour. Since previous work based on H/He-REOS.1 \citep{Fortney11, Nettelmann13} or the \citet{SCvH95} data \citep{Linder19} did reproduce the age of $\sim \SI{4.6}{Gyr}$, the shortening found here implies that Neptune seems to be incompatible with the known age of the solar system and appears to be brighter than predicted by adiabatic cooling. As is  shown in Sects. \ref{sec:H2OEOS} and \ref{sec:sol}, with the same set of assumptions the luminosity of Uranus is also not able to be reproduced. 
\subsection{Water equation of state}
\label{sec:H2OEOS}
\begin{figure}
  \resizebox{\hsize}{!}{\includegraphics{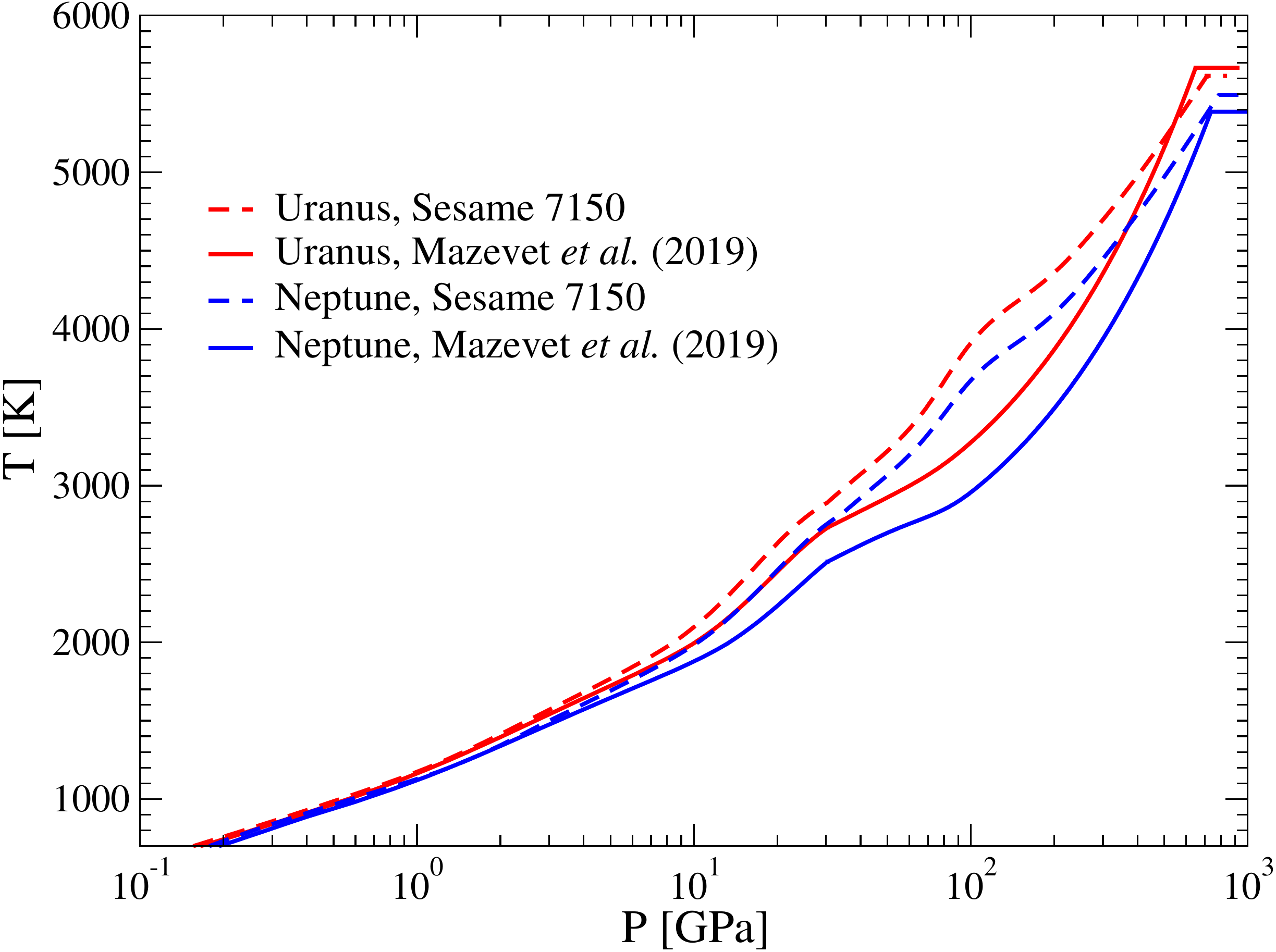}}
  \caption{Deep interior of Uranus (red) and Neptune (blue) with the heavy elements in the envelope treated via the Sesame water EOS (dashed) or the Mazevet EOS (solid), respectively.} \label{fig:UN_zoomed} 
\end{figure}

\begin{figure*}
    \centering
    \includegraphics[width=17cm]{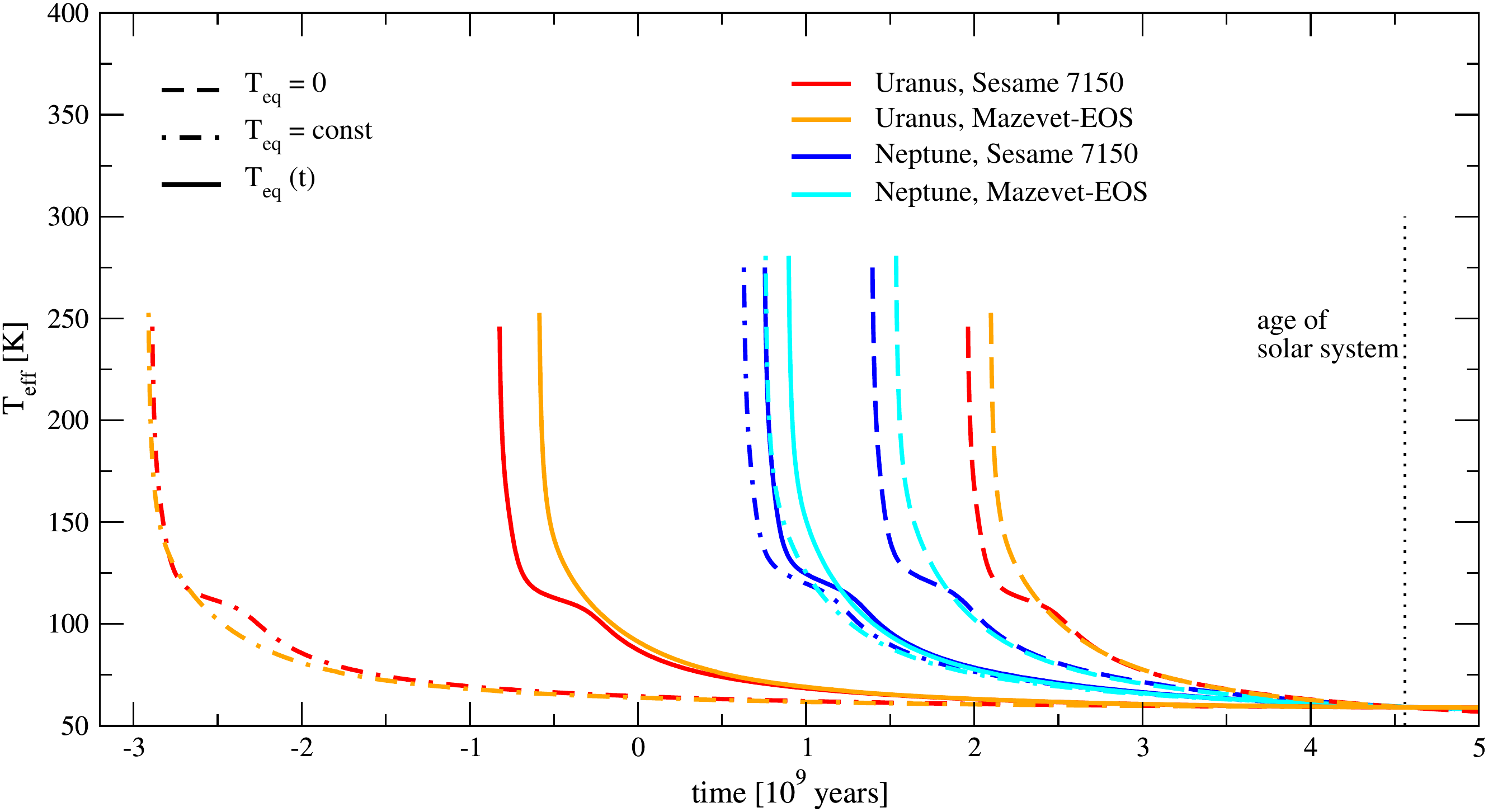}
    \caption{Thermal evolution of Uranus (red and orange) and Neptune (blue and cyan). Dashed lines represent no solar incident flux, dot-dashed constant solar irradiation, and solid lines linearly changing solar luminosity (see sect.~\ref{sec:sol}). Also shown are two different EOSs for the water in the envelope, SESAME 7150 \citep{Ree76} and \citet{Mazevet19}. }  \label{fig:main}
  \end{figure*}

In fig.~\ref{fig:UN_zoomed} we compare adiabatic $P-T$ relations using the Mazevet EOS and the Sesame EOS for water. These curves are taken from the structure models introduced in Table~\ref{tab:parameters}. For pressures larger than a few GPa, there are noticeable differences between the two models. Nevertheless, the two planets end up with similar values for the central temperature independently of water EOS, leading to a core temperature of about $T_\text{c} \approx \SI{5700}{\kelvin}$ for Uranus and $T_\text{c} \approx \SI{5500}{\kelvin}$ for Neptune.\\
Although it is  cooler, the Mazevet-EOS based adiabats require a larger ice mass fraction to match the planet radius (see table~\ref{tab:parameters}). A greater  heavy element content, or a smaller H/He mass fraction, typically reduces the energy budget of a planet and shortens the cooling time, see Fig.~\ref{fig:main}, where using the Mazevet EOS instead of Sesame leads to a further shortening of the cooling times by about $\sim 5\%$. The cooling curves using Sesame also feature an edge at $T_\text{eff} \sim \SI{120}{K}$ not seen in the Mazevet curves. This is caused by the behaviour of the individual adiabats'  in the deep interior of the planets at conditions of about $\SI{100}{GPa}$ and $\SI{5000}{K}$ -- ${7000}{K}$ and by the differences of the EOS in that regime. However, the effect on cooling time, while noticeable, is far less than that of other influences investigated here, such as H/He EOS or solar influx. Crucially, using the Mazevet EOS also produces cooling times that are too short for Neptune and too long for Uranus. 

\subsection{Influence of solar irradiation}
\label{sec:sol}
\begin{figure}
  \resizebox{\hsize}{!}{\includegraphics{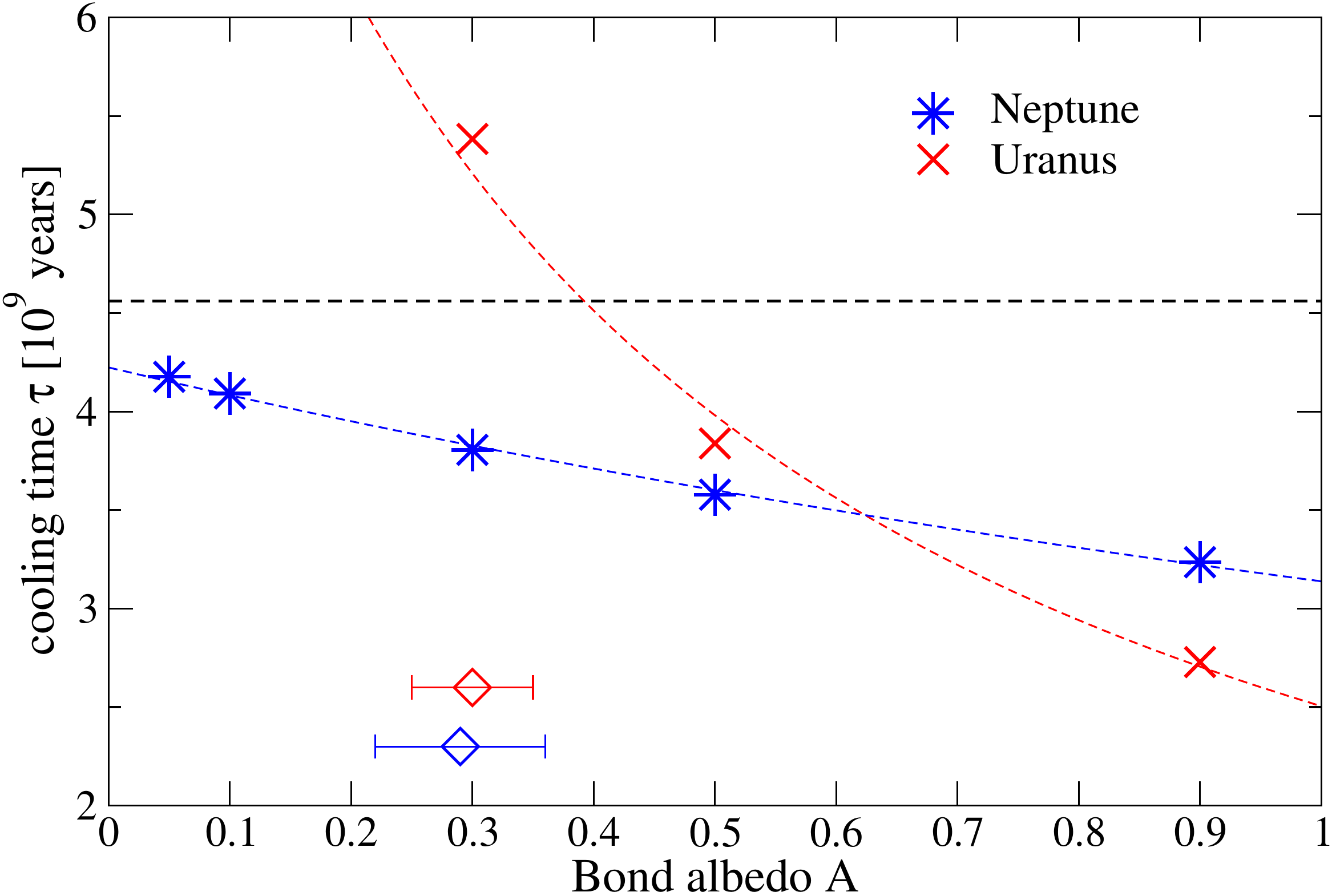}}
  \caption{Cooling times of Uranus and Neptune for different values of the Bond albedo. All calculations shown use Sesame EOS 7150 for water and assume a linearly changing solar luminosity with time. Diamonds indicate the observed Voyager values of $A$ \citep{Pearl90, Pearl91}. The thin dashed lines show the hyperbolic curve fit as a guide to the eye.} \label{fig:Albedo}
\end{figure}

An important part in calculating the evolution of a planet is the treatment of external energy being deposited into it. In our model this is expressed via the luminosity boundary condition equation \eqref{eq:B1}, specifically the quantity of the absorbed and re-emitted solar luminosity $L_\text{sol}$. It can be expressed as
\begin{equation}
  L_\text{sol} = 4 \pi R^2_\text{p} \sigma_\text{B} T_\text{eq}^4 ,
\end{equation}
with the equilibrium temperature
\begin{equation}
  T^4_\text{eq}=\frac{1}{4} (1-A) \left(\frac{R_*}{a} \right)^2 T^4_*.
\end{equation}
Here $R_*$ denotes the Sun's radius, $T_*$ its effective temperature, $a$ the planet's semi-major axis, and $A$ its Bond albedo. All our calculations assume the latter two parameters to be fixed over the planet's evolution. \\
We consider three cases: $T_\text{eq}=0$, $T_\text{eq}=\text{const.}$, and $T_\text{eq}(t)$, which accounts for the linear increase in the Sun's luminosity over time. Although the equilibrium temperatures of Uranus and Neptune are quite small with $T_\text{eq,U}=\SI{58.1 \pm 1.1}{K}$ and $T_\text{eq, N}=\SI{46.4 \pm 1.1}{K}$ \citep{Guillot05}, fig.~\ref{fig:main} shows a considerable influence on calculated cooling time. Entirely neglecting the solar irradiation ($T_\text{eq}=0$) produces very short cooling times for both Uranus and Neptune. Notably, in this case Uranus cools more quickly than Neptune. Fixing the Sun's luminosity to its present-day value over the entire evolution ($T_\text{eq}=\text{const.}$), as was done   by \citet{Fortney11} or \citet{Nettelmann13}, among others, gives significantly longer cooling times by about $\SI{1e9}{years}$ for Neptune and several billion years for Uranus. Cooling times for $T_\text{eq}(t)$, which is the most realistic case, fall   between these extremes.\\ 
The differences found when using the three approaches are vastly more pronounced for Uranus than for Neptune. This is due to the fact that Uranus' present-day effective temperature $T_\text{eff}=\SI{59.1\pm 0.3}{K}$ \citep{Guillot05} is very close to its equilibrium value, while Neptune is further away from that equilibrium with $T_\text{eff}=\SI{59.3 \pm 0.8}{K}$ \citep{Guillot05}. The slope of the cooling curve becomes very shallow as the planet approaches equilibrium, meaning that small changes in how that equilibrium is treated affect Uranus more than Neptune. This means that modifying assumptions such as the heavy element distribution or the spherical approximation have a stronger impact on Uranus than on Neptune so that cooling times presented here for Uranus have a rather large uncertainty. In addition, we find that the observational error bars of $T_\text{eff}$ induce an additional uncertainty of about $\sim \SI{0.5}{Gyr}$ for Uranus and $\sim \SI{0.3}{Gyr}$ for Neptune assuming adiabatic cooling.\\
The importance of an accurate observational grip on the solar influx can also be seen in fig.~\ref{fig:Albedo} where we plot the cooling time as a function of Bond albedo $A$. As before, we see a more drastic influence on Uranus' cooling than on Neptune's. In particular, a change in Uranus' albedo from $0.3$ to $0.4$ would bring its cooling time into agreement with the age of the solar system. For Neptune, however, the cooling time is shorter than the known age for all values of $A$. We evaluate this as further evidence that the brightness of Neptune is inconsistent with a fully adiabatic interior.
\section{Conclusions}
\label{sec:conclusion}
We have presented homogeneous thermal evolution calculations for Uranus and Neptune and have investigated the impact of solar energy influx, Bond albedo, and equations of state of hydrogen, helium, and water on the cooling behaviour. Our preferred models using REOS.3 EOSs for H and He and \citet{Mazevet19} for water, as well as time-dependent solar luminosity predict a cooling time of about $\tau_\text{U}=\SI{5.1}{Gyrs}$ for Uranus and $\tau_\text{N}=\SI{3.7}{Gyrs}$ for Neptune. This is notably shorter than previous results and means that neither planet's present-day luminosity can be explained by adiabatic models. Our results therefore confirm that a more sophisticated approach is required than the fully adiabatic assumption. A non-adiabatic deep interior region has been suggested by \citet{Hubbard95} to explain the faintness of Uranus, and could also explain the magnetic field morphology of the ice giants \citep{Stanley04, Stanley06};  the assumption of a sharp temperature gradient at a layer boundary within Uranus' envelope was shown by \citet{Nettelmann16} to be able to reproduce Uranus' correct age while also being compatible with gravitational field measurements. Moreover, physical processes like sedimentation or upwelling of a major constituent in the HCNO system could also have a large influence on the thermal evolution of the ice giants. For instance, it is possible, that demixing of hydrogen and water occurs at conditions relevant for the ice giant envelope \citep{Bali13}, although ab initio simulations suggest solubility of water and hydrogen under Uranus and Neptune interior conditions \citep{Soubiran15}. Deeper in the interior at pressures of about $\SI{1.5}{Mbar}$, carbon-hydrates have been experimentally found to separate into diamond, which may sink to the core, and hydrogen, which may rise upward \citep{Kraus17}.  In the same vein, while first-principle calculations of 1:1 water-ammonia mixtures have suggested the presence of a stable superionic phase of H$_2$O and NH$_3$ \citep{Bethkenhagen15}, additional crystal structure searches have revealed that ammonia, when mixed with water in solar proportions, may  preferentially connect to protons donated by a superionic H$_2$O-ice lattice and thus perhaps form an ocean of light ices on top of the water-rich deep interior \citep{Robinson17}. Such a layer could lead to stable stratification across the region identified as a layer boundary in conventional three-layer structure models. All these processes can have a strong impact on the planetary structure and thermal evolution.\\
Additionally, uncertainties on the solar flux due to the present Bond albedo, which we assume to be constant over time, can have considerable influence on the planet's cooling behaviour, which is especially pronounced for Uranus, because it is very close to being in thermal equilibrium with the Sun. This is problematic because the current albedo values of the ice giants have relatively large observational uncertainties, which could even be larger than the currently accepted error bars. We note that Cassini observations of the Jovian atmosphere recently led to a significantly higher Bond Albedo of \num{0.50} compared to the Voyager value of \num{0.34} \citep{Li18}.\\
Improved observational constraints on the present values of albedo and effective temperature are thus clearly much needed also for the ice giants. Observational constraints for earlier times when the atmosphere was warmer will have to wait for albedo measurements of Neptune-like exoplanets around young Sun-like stars.\\
\section{Acknowledgements}
We thank the referee for helpful comments for this publication. We thank Martin French for his help in adapting the Mazevet EOS for our use, in particular for applying the Maxwell construction, and for providing fruitful discussion of the results. This project was funded by the German Research Foundation DFG as part of the Research Group FOR-2440 \glqq Matter Under Planetary Interior Conditions \grqq. We also acknowledge the hospitality of the International Space Science Institute in Bern.

\bibliographystyle{aa}
\bibliography{aa36378-19.bib} 
\end{document}